\documentclass[cameraready]{Interspeech}
\usepackage{multirow}
\title{ImKWS: Test-Time Adaptation for Keyword Spotting with Class Imbalance}

\author[affiliation={1}, equalcontribution]{Hanyu}{Ding}
\author[affiliation={2}, equalcontribution]{Yang}{Xiao}
\author[affiliation={2}]{Jiaheng}{Dong}
\author[affiliation={2}]{Ting}{Dang}

\address{
    $^1$ Jiangsu University \\
    $^2$ The University of Melbourne
}

\email{2212308044@stmail.ujs.edu.cn, yang.xiao.1@student.unimelb.edu.au}


\usepackage{comment}
\usepackage{soul,xcolor}

\usepackage{hyperref}
\usepackage{cite}
\usepackage[table]{xcolor}
\definecolor{myblue}{RGB}{200,220,235}

\begin{document}

\maketitle

% The abstract here must exactly match the abstract entered into the paper submission system (max 1000 characters, no citations).
\begin{abstract}
% Keyword spotting (KWS) models degrade under distribution shift such as environmental noise. We improve test-time adaptation (TTA) by selecting reliable pseudolabels via entropy and augmentation-consistency (PLPD), then reweighting samples and applying consistency regularization. On ESC-50 and MS-SNSD noise settings, our method consistently outperforms unadapted and five TTA baselines (TBN, TENT, SAR, EATA, AdaKWS) in accuracy, with the largest gains at low SNR. Ablations show that decoupled entropy minimization and consistency loss each contribute.
Keyword spotting (KWS) identifies words for voice assistants, but environmental noise frequently reduces accuracy. Standard adaptation fixes this issue and strictly requires original or labeled audio. Test-time adaptation (TTA) solves this data constraint using only unlabeled test audio. However, current methods fail to handle the severe imbalance between rare keywords and frequent background sounds. Consequently, standard entropy minimization becomes overconfident and heavily biased toward the frequent background class. To overcome this problem, we propose a TTA method named ImKWS. Our approach splits the entropy process into a reward branch and a penalty branch with separate update strengths. Furthermore, we enforce consistency across multiple audio transformations to ensure stable model updates. Experiments on the Google Speech Commands dataset indicate ImKWS achieves reliable adaptation in realistic imbalanced scenarios. The code is available on GitHub~\footnote{\url{https://github.com/dhyzy123/ImKWS}}.
\end{abstract}

\vspace{-2mm}
\section{Introduction}
\label{sec:intro}

% Keyword spotting (KWS) systems are often deployed in conditions that differ from training (e.g., environmental noise), leading to distribution shift and degraded accuracy. Test-time adaptation (TTA) updates the model on unlabeled test batches to reduce this gap, but standard entropy-minimization approaches are sensitive to erroneous pseudolabels and can accumulate errors.

% We propose to improve TTA for KWS under environmental noise by (1) \textbf{two-stage filtering}---selecting samples first by low entropy, then by high augmentation-consistency (PLPD)---so that only confident, stable pseudolabels drive updates; (2) \textbf{reweighting} selected samples by entropy and PLPD; and (3) \textbf{consistency regularization} on spec-augmented views. We combine decoupled entropy minimization with a symmetric consistency loss. On ESC-50 and MS-SNSD across noise settings and SNRs, our method achieves the best or tied-best accuracy in all reported conditions, with the largest gains at low SNR. Ablations confirm that both decoupled entropy and consistency contribute.

Keyword spotting (KWS) \cite{lopez2021deep} aims to detect specific keywords from continuous speech streams. It is widely used in applications such as smart device control, voice assistants, and voice search \cite{zhang2017hello,kim2021broadcasted,xiao2022rainbow}. The primary goal of KWS \cite{ding2025keyword,peng2025dark,xiao2025analytickws,luo2026survey} is to achieve better detection performance while satisfying the constraints of low power consumption and limited computational resources.

Despite these efforts to optimize efficiency, these systems frequently fail when placed into unpredictable real-world acoustic environments. Specifically, sudden background noise alters the audio patterns compared to the initial training data. When the test data distribution shifts significantly, detection performance typically degrades. Traditional solutions usually rely on supervised fine-tuning \cite{wu2020domain,lopez2021novel,yang2025cross,ozay2024joint,xiao2025dg,xiao2026adapting} or unsupervised domain adaptation \cite{ganin2015unsupervised,hou2019domain,liu2023te,lim2023joint}, but they require access to a small amount of labeled target-domain data or the original source-domain data. However, in dynamic, real-world deployments, anticipating sudden shifts to collect and label target data on-the-fly is practically impossible. Furthermore, continuously storing or transmitting source data raises severe privacy and memory concerns for resource-constrained devices. 
% \jd{Better not mention "edge", just resource-constrained devices}

% \jd{If previously we say "model fail in unpredictable real-world environemnt", which seems the target data is not practical to be labeled instead of labeling costs?}

Recently, test-time adaptation (TTA) has emerged as a strong solution \cite{liang2025comprehensive,wang2021tent,niu2023towards,niu2022efficient,schneider2020improving,dong2025ebats,dang2026test,dong2026test,shi2026tta}. This approach adjusts models to unseen environments at inference time using only unlabeled target data, without requiring access to source data. Not long ago, researchers tailored this concept for large speech foundation models. For instance, methods like SUTA \cite{lin2022listen} update feature and normalization layers and incorporate temperature smoothing in automatic speech recognition (ASR).
% ~\jd{feature and normalization layers?} \textcolor{red}{[already modified]} 
% Following this, approaches such as SGEM \cite{kim2023sgem} and CEA \cite{liu2024advancing} design advanced loss functions to handle sequence-level and frame-level uncertainty, respectively. Furthermore, frameworks such as DSUTA \cite{lin2024continual} and AWMC \cite{lee2023awmc} introduce multi-branch architectures to stabilize continual adaptation across sequential audio recordings. Additionally, methods like E-BATS \cite{dong2025ebats} achieve a practical balance between adaptation effectiveness and memory efficiency. While these methods demonstrate promising performance on tasks such as automatic speech recognition that operate on complete utterances with rich temporal context, keyword spotting presents distinct challenges. \jd{Double check if this is what u mean, previously u state it as the model differences, large speech foundation models V.S. KWS models, but below it seems more like the task differences, i have adjusted this, check if it is correct in KWS}

In contrast, lightweight KWS models relies on short audio windows with limited contextual information 
% \jd{Updated, have a look}
, making them prone to error accumulation from unreliable, high-entropy samples during adaptation. For example, a recent study named AdaKWS \cite{xiao2025adakws} successfully adapts small KWS models through selective entropy minimization (EM). However, this method ignores a critical reality \cite{ng2023small}. The ratio between keyword samples and background sounds is extremely imbalanced in continuous speech. During entropy minimization, most of the updates are driven by background segments. Over time, the model becomes increasingly confident in predicting the background class, gradually shifting the decision boundary and making it harder to detect rare keyword events.

% When a model adapts to this uneven data, standard entropy minimization becomes blindly overconfident. It creates a heavy bias toward the frequent background sounds, which causes poor adaptation and severe performance drops. 

% \jd{I think soften this will better, something like: During entropy minimization, most of the updates are driven by background segments. Over time, the model becomes increasingly confident in predicting the background class, gradually shifting the decision boundary and making it harder to detect rare keyword events.}

To address this data imbalance challenges in real-world KWS,
% \jd{what are "these challenges" refers to? It seems the only challenge is data imbalance problem?} \textcolor{red}{[lightweigh and imbalance ?]}
we propose ImKWS, a novel method designed to mitigate severe class imbalance in dynamic environments. While standard entropy minimization is widely used for TTA, it can be overly confident when faced with skewed data streams. To address this, we decouple the standard entropy objective into an independent reward branch and a penalty branch. By using adjustable parameters to control the update strength of each branch separately, our model maintains sensitivity to the rare keyword class while remaining strictly conservative toward frequent background sounds. In addition, we apply multi-view data augmentation to ensure consistent predictions across different audio transformations. This consistency suppresses severe gradient fluctuations caused by the imbalanced data.
% \jd{If this is also the challenge, put them on previous paragraphs.} \textcolor{red}{[This issue is brought about by dem. I made a minor adjustment in section 2.2.2, but I'm unsure if the wording is correct.]} 
Extensive experiments on the Google Speech Commands dataset under various noise conditions confirm that ImKWS delivers highly robust adaptation. To our knowledge, this is the first study to explore TTA for KWS in realistic imbalanced scenarios.

% To overcome these challenges, we propose a new method named ImKWS. This approach specifically targets the severe class imbalance in dynamic environments. Building on existing foundations, we divide the standard entropy calculation into a reward branch and a penalty branch. Adjustable parameters separately control the update strength of each branch. Consequently, our model reacts strongly to the rare keyword class but remains conservative toward the frequent background sounds. In addition, we apply multiple view data augmentation to ensure consistent predictions across different audio transformations. This consistency suppresses severe gradient fluctuations caused by the uneven data. We evaluated our approach on the Google Speech Commands dataset under various noise conditions. The experiments confirm that ImKWS delivers highly robust adaptation. To our knowledge, this is the first study to explore test time adaptation for keyword spotting in realistic imbalanced scenarios.

\vspace{-2mm}
\section{Method}
\label{sec:method}

\begin{figure*}[t!]
    \centering
    \includegraphics[width=0.618\textwidth]{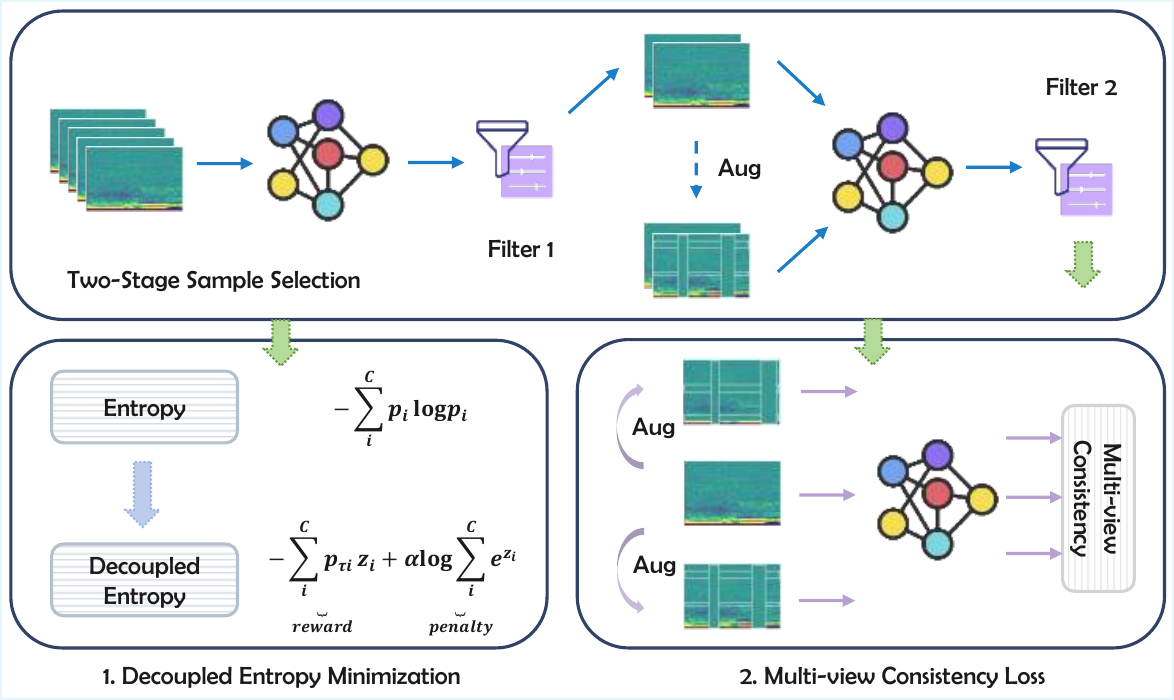}
    \vspace{-4mm}
    \caption{Overview of the Proposed ImKWS Method, including Two-stage Sample Selection, Decoupled Entropy Minimization and Multi-view Consistency Loss.}
    \label{fig:tta-kws}
    \vspace{-8mm}
\end{figure*}

% We assume a pretrained KWS model and a stream of test batches without labels. Only batch normalization (BN) parameters are updated at test time. For each batch we (1) compute decoupled entropy per sample and keep samples with entropy below a threshold; (2) for those samples, obtain augmented views (e.g., spec augmentation) and define PLPD as the drop in confidence of the pseudolabel class on the augmented view; keep samples with PLPD above a small threshold (augmentation-consistent); (3) reweight the retained samples by a coefficient that favours low entropy and high PLPD; (4) minimise the reweighted entropy plus a symmetric cross-entropy consistency term between the original and two augmented outputs. Loss is backpropagated only through the selected samples and only into BN parameters. Hyperparameters: entropy margin 0.4, PLPD threshold 0.05, consistency weight 0.7; optimizer SGD, lr $1\times10^{-4}$, batch 128.

\subsection{Problem Formulation}
% Adaptation aims to improve model generalization from the training domain to the testing domain. Specifically, we define a trained KWS model as $\mathcal{M}_{\theta}$. This model is trained on a source domain $D_{\mathrm{train}}=\left(x_i^{\mathrm{train}},\,y_i^{\mathrm{train}}\right)_{i=1}^{N_{\mathrm{train}}}$ and parameterized by pre-trained weights $\theta$, where $x^{\mathrm{train}}\in\mathcal{X}^{\mathrm{train}}$ and $y^{\mathrm{train}}\in\mathcal{Y}$. The main goal of TTA is to adapt $\mathcal{M}_{\theta}$ successfully using only target domain test data $D_{\mathrm{test}}=\left(x_i^{\mathrm{test}},\,y_i^{\mathrm{test}}\right)_{i=1}^{N_{\mathrm{test}}}$, where $x^{\mathrm{test}}\in\mathcal{X}^{\mathrm{test}}$ and $y^{\mathrm{test}}\in\mathcal{Y}$. Unlike fine-tuning and UDA, TTA does not access $D_{\mathrm{train}}$ or the test labels $y^{\mathrm{test}}$ during the adaptation process. Consequently, this approach can reduce both data dependency and computational costs.

Let $\mathcal{M}_{\theta}$ be a KWS model parameterized by pre-trained source weights $\theta$. Test-time adaptation (TTA) aims to adapt $\mathcal{M}_{\theta}$ to a new target domain using only an unlabeled test stream $D_{\mathrm{test}}=\left(x_i\right)_{i=1}^{N_{\mathrm{test}}}$. Unlike standard fine-tuning or unsupervised domain adaptation, TTA operates strictly without access to the original source training data or the target labels $y_i$. %This constraint eliminates heavy data dependencies and privacy concerns, making it highly suitable for spoken keyword spotting models deployed on resource-constrained edge devices.

\subsection{Proposed ImKWS Method}
\subsubsection{Decoupled Entropy Minimization}
Existing TTA methods for KWS (e.g., AdaKWS~\cite{xiao2025adakws}) predominantly rely on standard Entropy Minimization (EM) for adaptation. However, in continuous KWS streams where background samples dominate, EM may bias updates toward the majority class, potentially affecting keyword detection performance.

To address this issue, we propose Decoupled Entropy Minimization (DEM). Standard EM typically forces the model to make confident predictions by minimizing the Shannon entropy of the output distribution. For an input sample $x$, let $z = \mathcal{M}_{\theta}(x) \in \mathbb{R}^C$ denote the output logits, and let $p_i = \frac{e^{z_i}}{\sum_{k=1}^{C} e^{z_k}}$ be the softmax probability for class $i$, where $C$ is the total number of classes. We also decompose the standard conditional entropy into a reward term and a penalty term: \vspace{-3mm}
\begin{equation}  \resizebox{0.9\hsize}{!}{$
\mathcal{L}_{ent}(x) = - \sum_{i=1}^{C} p_{i} \log p_{i} = \underbrace{- \sum_{i=1}^{C} p_{i}z_{i}}_{T(z) \text{ (Reward)}} + \underbrace{\log \sum_{i=1}^{C} e^{z_{i}}}_{Q(z) \text{ (Penalty)}}.$}
\end{equation}

\vspace{-2mm} 
A vanilla decomposition is insufficient for imbalanced data streams. Therefore, we reshape each branch independently to maintain sensitivity to rare keywords while explicitly throttling majority-class overconfidence.

\textbf{Reward Branch.} To ensure that the adaptation signal for minority keywords remains stable, we introduce a temperature parameter $\tau$ to the reward branch to control the sharpness of the predicted distribution: \vspace{-2mm}
\begin{equation}
  \resizebox{0.9\hsize}{!}{$
T_{\tau}(z) = - \sum_{i=1}^{C} p_{\tau i}z_{i}, \quad \text{where } p_{\tau i} = \frac{e^{z_{i}/\tau}}{\sum_{k=1}^{C} e^{z_{k}/\tau}}.$}
\end{equation}

\vspace{-2mm}

\textbf{Penalty Branch.} The core mechanism for mitigating class imbalance lies in adjusting the penalty term $Q(z)$. Standard EM aggressively pushes the maximum logit toward $+\infty$ and non-target logits toward $-\infty$, causing severe overconfidence. To counteract this, we introduce a tunable scaling factor $\alpha$ as:

\vspace{-3mm}
\begin{equation}
Q_{\alpha}(z) = \alpha \log \sum_{i=1}^{C} e^{z_{i}}.
\end{equation}

\vspace{-1mm}
To rigorously understand how $\alpha$ rectifies majority-class bias, we analyze the gradient of the formulated DEM loss ($\mathcal{L}_{dem} = T_{1.0}(z) + Q_{\alpha}(z)$) with respect to a specific logit $z_{j}$ as the following equation:
\vspace{-2mm}
\begin{equation}
\frac{\partial \mathcal{L}_{dem}}{\partial z_{j}} = p_{j}(z) \left( \sum_{i=1}^{C} p_{i}(z)z_{i} - z_{j} - (1 - \alpha) \right).
\end{equation}

\vspace{-2mm}

% For a non-target class $j \neq \arg\max z_{i}$, the logit $z_{j}$ is relatively small, making the term $(\sum p_{i}z_{i} - z_{j})$ positive\td{and the term $(\sum_i p_i z_i - z_j)$ tends to be positive. You cannot make sure it is always positive given it is the expectation minus a logit value, right?}. In standard EM ($\alpha = 1.0$), this results in a strictly positive gradient\td{is it a strictly positive value? the expectation minus a non target logit can also be negative? z=[10,9.8,1] }, which aggressively suppresses $z_{j}$ toward $-\infty$ during gradient descent\td{also for gradient, it is not pushing it to infinity right? It just decreases the gradient directions. Entropy minimization does not drive logits to infinity. It just reshapes the logit distribution by increasing confidence in dominant classes and suppressing non-dominant ones. Your method seems more accurate to say that it induces a continuous decrease in non-target logits, as a reduction in the strength of the gradient signal rather than suggesting a infinite directional effect.}. By setting $\alpha < 1.0$ (e.g., $\alpha = 0.8$), we subtract a strictly positive margin $(1 - \alpha)$ from the gradient. This explicitly \textit{weakens} the downward push on non-target logits. Consequently, this throttling mechanism prevents the network from aggressively driving non-target predictions to zero, serving as a direct regularizer against the overconfident one-hot predictions typical of the majority background class.

For a non-target class $j \neq \arg\max z_{i}$, the logit $z_{j}$ is relatively small, making the term $(\sum p_{i}z_{i} - z_{j})$ tend to be positive for the majority of non-target classes. In standard EM ($\alpha = 1.0$), this results in a positive gradient that induces a continuous decrease in non-target logits and reshapes the distribution by persistently increasing confidence in dominant classes while suppressing non-dominant ones. By setting $\alpha < 1.0$ (e.g., $\alpha = 0.8$), we subtract a positive margin $(1 - \alpha)$ from the gradient, reducing the strength of the gradient signal that pushes down non-target logits. Consequently, this mechanism prevents the network from aggressively driving non-target predictions to zero, serving as a regularizer against the overconfident one-hot predictions typical of the majority background class.

\subsubsection{Multi-view Consistency Loss}
% Moreover, while DEM mitigates EM-induced overconfidence, imbalanced testing streams remain highly vulnerable to acoustic perturbations.
% Although DEM alleviates the overconfidence issue caused by the EM algorithm, it increases gradient norm fluctuation~\jd{Why it increase gradient norm fluctuation,?}, making the imbalanced test stream more susceptible to acoustic perturbations.
DEM mitigates the overconfidence issue caused by EM because it actively constrains the majority-class predictions and prevents minority-class gradient signals from being drowned out by majority-class predictions. However, DEM also inadvertently amplifies the relative impact of individual noisy samples and increases gradient norm fluctuations.
% ~\jd{Why?} \textcolor{red}{[I'm not sure. already modified]} 
To ensure gradient stability, we enforce consistency regularization across multiple augmented views (e.g., time and frequency masking). Given an input $x$ and two augmented views $\tilde{x}$ and $\hat{x}$, the consistency loss is defined as:
\vspace{-1mm}
\begin{equation}
\mathcal{L}_{consist}(x, \tilde{x}, \hat{x}) = \mathcal{L}_{sce}(x, \tilde{x}) + \mathcal{L}_{sce}(x, \hat{x}).
\end{equation}

\vspace{-1mm}
Here, $\mathcal{L}_{sce}(x, \tilde{x})$ denotes the symmetric cross-entropy \cite{wang2019symmetric}, and $\mathcal{L}_{sce}(x, \hat{x})$ is defined analogously:
% \vspace{-1mm}
%   \begin{equation}
%   \resizebox{0.9\hsize}{!}{$
% \mathcal{L}_{sce}(x, x') = - \frac{1}{2} \left( \sum_{i=1}^{C} p_{i}(x) \log p_{i}(x') + \sum_{i=1}^{C} p_{i}(x') \log p_{i}(x) \right),$}
% \end{equation}

\vspace{-2mm}
  \begin{equation}
  \resizebox{0.9\hsize}{!}{$
\mathcal{L}_{sce}(x, \tilde{x}) = - \frac{1}{2} \left( \sum_{i=1}^{C} p_{i}(z) \log p_{i}(\tilde{z}) + \sum_{i=1}^{C} p_{i}(\tilde{z}) \log p_{i}(z) \right),$}
\end{equation}  
% \vspace{-1mm}
where $z$ and $\tilde{z}$ denote the logits produced by the model for inputs $x$ and $\tilde{x}$, respectively. This strategy is motivated by the strong tolerance and robustness of $\mathcal{L}_{sce}$ to label noise; consequently, even when predictions are biased or unreliable, it can still provide a more stable training signal. Moreover, it facilitates better learning across both easy and hard classes to improve overall performance.

% \jd{Why need a new symbol for augmented view?} \textcolor{red}{[Since two enhanced symbols $\tilde{x}$ and $\hat{x}$ are used above, I'm unsure which one is appropriate. already modified]} 

% This choice is motivated by the tolerance of $\mathcal{L}_{sce}$ to label noise, which is especially valuable where predictions can be inaccurate.

% \jd{What is "self-training" here, it seems this word is jumped out suddenly.} \textcolor{red}{[already modified]}

% To prevent highly uncertain or noisy background segments from dominating the gradients, we formulate $t$ as a sample-specific confidence weight, defined exactly as $t = \max_{i}(p_{i}(x))$.

\subsubsection{Overall: Two-Stage Sample Selection and Objective}
As illustrated in Figure \ref{fig:tta-kws}, ImKWS employs a robust two-stage sample selection strategy based on AdaKWS to filter the imbalanced stream before enforcing the adaptation losses.

\textbf{Two-Stage Sample Selection.} The sample selection consists of Selective Entropy Minimization (actually implemented via DEM here) and the pseudo-keyword consistency (PKC) based resampling scheme. The test sample $x$ is selected if and only if it meets two predefined thresholds $\tau_{dem}$ and $\tau_{pkc}$:
\vspace{-1mm}
\begin{equation}
\resizebox{0.9\hsize}{!}{$x_{s}=\left\{x \mid \mathcal{L}_{dem}(x)<\tau_{dem},\ \mathcal{L}_{pkc}(x, x')>\tau_{pkc}\right\},$}
\end{equation}
% \vspace{-1mm}
where $\mathcal{L}_{pkc}(x,x') = p(x)_c - p(x')_c$, and $p(x)_c$ and $p(x')_c$ are the model's confidence scores for pseudo-label $c$ on the original input $x$ and the transformed input $x'$, respectively. Let $x_{s}$ denote the subset of samples that pass this criterion, with $\tilde{x}_{s}$ and $\hat{x}_{s}$ representing their corresponding augmented views. 
% ~\jd{numerical settings could be moved to experimental setup?} \textcolor{red}{[already modified]}

% \begin{equation}
% \text{Condition: } \max_{i}(p_{i}(x)) > \theta_{pkc},
% \end{equation}
% where we empirically set $\tau_{dem}=0.4$ and $\tau_{pkc}=0.05$, and define $\mathcal{L}_{pkc}(x,x') = p(x)_c - p(x')_c$, with $p(x)_c$ and $p(x')_c$ representing the model confidence for pseudo-label $c$ on the original input $x$ and the transformed input $x'$, respectively. Let $x_{s}$ denote the subset of samples that pass this criterion, with $\tilde{x}_{s}$ and $\hat{x}_{s}$ representing their corresponding augmented views.

\textbf{Objective.} For the samples that pass two-stage sample selection, the overall objective function for ImKWS is a weighted combination of DEM and the multi-view consistency loss:
\vspace{-1mm}
\begin{equation}
\resizebox{0.9\hsize}{!}{$
\mathcal{L}_{total} = w(x_s) \cdot \mathcal{L}_{dem}(x_{s}) + \lambda \cdot \mathcal{L}_{consist}(x_{s}, \tilde{x}_{s}, \hat{x}_{s}).$}
\end{equation}

\vspace{-1mm}
% Here, the sample weight is computed as 
% $w(x_{s})=\frac{1}{\exp\{\mathcal{L}_{dem}(x_{s})-\sigma\}}+\frac{1}{\exp\{-\mathcal{L}_{pkc}((x_{s},(x_{s}')\}}$, where $\mathcal{L}_{\mathrm{dem}}(x)$ denotes the entropy loss, $\mathcal{L}_{\mathrm{pkc}}(x,x')$ represents the PKC score, and $\sigma$ is a normalization factor set to $0.5$. For the consistency term, we explicitly set its coefficient $\lambda$ to $0.7$.

Here, the sample weight $w(x)$ is computed based on the decoupled entropy loss $\mathcal{L}_{dem}(x)$ and the PKC score $\mathcal{L}_{pkc}(x, x')$, with $\sigma$ denoting the normalization factor:
\vspace{-1mm}
\begin{equation}
\resizebox{0.8\hsize}{!}{$\displaystyle
w(x)=\frac{1}{\exp\{\mathcal{L}_{dem}(x)-\sigma\}}+\frac{1}{\exp\{-\mathcal{L}_{pkc}(x,x')\}},$}
\end{equation}
and $\lambda$, which is a constant, denotes the consistency coefficient.
% Here, the sample weight $w(x)$ is computed based on the entropy loss $\mathcal{L}_{dem}(x)$ and the PKC score $\mathcal{L}_{pkc}(x, x')$. The normalization factor $\sigma$ is set to $0.5$. Specifically, the weight is defined as:
% and the consistency coefficient is set to $\lambda = 0.7$.
% \jd{Same comments here, move to experimental setup?} \textcolor{red}{[already modified]}

\section{Experimental Setup}

% We use ESC-50 and MS-SNSD environmental noise. We compare unadapted, TBN, TENT, SAR, EATA, AdaKWS, and ours. Metrics: Accuracy (Acc), Balanced Accuracy (BAcc), Weighted/Micro/Macro F1. The model checkpoint is selected by balanced accuracy on a held-out validation set.

\textbf{Dataset.} We adopt the 12-class Google Speech Commands v2 dataset \cite{warden2018speech} sampled at 16 kHz, split into training (80\%), validation (10\%), and test (10\%) sets. To construct a targeted 4-class classification task, we select ``yes'', ``up'', and ``stop'' as positive keywords and merge the remaining nine categories into a single non-keyword class. The source distribution is preserved for training and validation. For testing, the data is resampled to simulate severe class imbalance, varying the keyword-to-non-keyword ratio from 1:4 to 1:8. To simulate acoustic distribution shifts, we corrupt the test audio with multi-source noise from ESC-50 \cite{piczak2015esc} and five types of real-world single-source noise from the MS-SNSD \cite{reddy2019scalable} test set, evaluating across three signal-to-noise ratios (SNRs).

\textbf{Baselines.} To measure our adaptation performance, we compare our method with several TTA baselines. First, Test-Time Normalization (TBN) \cite{schneider2020improving} updates the batch normalization statistics using target data during testing. Next, Tent \cite{wang2021tent} adapts the affine parameters of normalization layers through standard entropy minimization. Additionally, ETA \cite{niu2022efficient} filters out high-entropy samples to prevent noisy gradients from degrading performance. Similarly, SAR \cite{niu2023towards} combines sharpness-aware optimization with reliable entropy minimization to handle practical problems like small batch sizes and online label shifts. Finally, AdaKWS \cite{xiao2025adakws} uses selective entropy minimization and a specific sample selection strategy to resolve sensitivity issues in KWS tasks.
% \jd{Add citations?} \textcolor{red}{[already modified]}

% \textbf{Implementation details:} To evaluate these methods, we employ BC-ResNet-3 as the network backbone. This architecture is a lightweight CNN designed specifically for on-device keyword spotting. During speech preprocessing, we extract 40-dimensional MFCC features as model inputs with a 160 ms hop length. For our ImKWS method, we set the parameters to $\tau=1.0$, $\alpha=0.8$, and $\lambda=0.7$. Furthermore, we apply two time-masks with a maximum length of 20 and two frequency-masks with a maximum length of 5 to augment the input features.

\textbf{Implementation details.} To evaluate the methods, we employ BC-ResNet-3 \cite{kim2021broadcasted} 
% \jd{could add a footnote like huggingface link here?} \textcolor{red}{[already modified]} 
as the network backbone, which is a lightweight convolutional neural network designed specifically for on-device KWS. During speech preprocessing, we extract 40-dimensional Mel-frequency cepstral coefficients as inputs with a 160 ms hop length. To augment the features for multi-view consistency, we apply two time-masks with a maximum length of 20 and two frequency-masks with a maximum length of 5. The adaptation is performed 
% in an online, continual streaming fashion 
with only a single pass over the test data. 
During this process, we update the affine parameters of the batch normalization layers. The optimization is conducted using the SGD optimizer with a learning rate of $1 \times 10^{-4}$ and a test-time batch size of 128. For the proposed ImKWS method, the hyperparameters are selected through grid search as follows: $\tau=1.0$, $\alpha=0.8$, and $\lambda=1.0$. For the two-stage sample selection strategy, we set $\tau_{dem}=0.4$, $\tau_{pkc}=0.05$, and $\sigma=0.5$.

% \textbf{Metrics:} To assess the detection performance comprehensively, we calculate the macro and micro averaged F1 scores. A higher score indicates better detection capability. Specifically, we choose these metrics because they accurately reflect the model's performance under the severe class imbalance present in our dataset.

\textbf{Metrics.} We report both macro- and micro-averaged F1 scores to evaluate performance under severe class imbalance. We mainly focus on macro F1, as it prevents the dominant non-keyword class from masking minority keyword recognition. 
\begin{table}[t]
\centering

\caption{Test-time adaptation results (macro F1 \& micro F1, \%) under ESC-50 and MS-SNSD environmental noises with keyword:non-keyword ratio = 1:8. Best per column is in \textbf{bold}.}
\vspace{-2mm}
\label{tab:1}
\setlength{\tabcolsep}{6pt}
\renewcommand{\arraystretch}{0.9}
\scalebox{0.8}{
\begin{tabular}{l l c|c|c}
\toprule
\multirow{2.5}{*}{\textbf{Dataset}} & \multirow{2.5}{*}{\textbf{Methods}} &
\multicolumn{3}{c}{\textbf{SNR (dB)}} \\
\cmidrule(lr){3-5}
& & $-10$ & $0$ & $10$ \\
\midrule

\multirow{7}{*}{\textbf{ESC-50}}

& Unadapted & 61.87 / 91.32 & 74.06 / 93.46 & 81.91 / 95.10 \\
& TBN       & 69.14 / 89.83 & 77.41 / 92.65 & 83.15 / 94.56 \\
& Tent      & 68.99 / 89.83 & 77.32 / 92.66 & 82.86 / 94.44 \\
& SAR       & 69.35 / 89.95 & 77.14 / 92.60 & 82.80 / 94.46 \\
& ETA       & 69.29 / 89.88 & 77.27 / 92.62 & 82.66 / 94.43 \\
& AdaKWS    & 69.68 / 90.25 & 77.55 / 92.72 & 82.89 / 94.47 \\
& ImKWS     & \textbf{70.91 / 91.20} & \textbf{78.98 / 93.57} & \textbf{84.51 / 95.23} \\
\midrule

% \multirow{7}{*}{\rotatebox[origin=c]{90}{\textbf{MS-SNSD}}}
\multirow{7}{*}{\textbf{MS-SNSD}}

& Unadapted & 61.33 / 90.75 & 73.69 / 92.88 & 80.44 / 94.65 \\
& TBN       & 66.66 / 89.54 & 74.43 / 91.85 & 79.80 / 93.55 \\
& Tent      & 67.06 / 89.98 & 74.86 / 92.13 & 79.72 / 93.54 \\
& SAR       & 66.25 / 89.46 & 74.06 / 91.81 & 79.63 / 93.53 \\
& ETA       & 66.52 / 89.58 & 74.61 / 92.01 & 79.95 / 93.61 \\
& AdaKWS    & 66.95 / 89.95 & 74.30 / 91.97 & 79.96 / 93.63 \\
& ImKWS     & \textbf{69.91 / 91.82} & \textbf{76.49 / 93.13} & \textbf{81.46 / 94.43} \\
\bottomrule
\end{tabular}}
\vspace{-18pt}
\end{table}

% -------- Table 2: Ratio --------
\newcommand{\pmS}[1]{\raisebox{0.15ex}{\scriptsize$\pm$#1}}
\begin{table*}[t]
\centering
\caption{Macro F1 \& Micro F1 (\%) at $-10$~dB across noise settings with keyword-to-non-keyword ratios ranging from 1:4 to 1:8.}
\vspace{-3mm}
\label{tab:3}
\setlength{\tabcolsep}{7pt}
\renewcommand{\arraystretch}{0.8}
\scalebox{0.8}{
\begin{tabular}{l l c|c|c|c|c}
\toprule
\multirow{2.5}{*}{\textbf{Dataset}} & \multirow{2.5}{*}{\textbf{Methods}} & \multicolumn{5}{c}{\textbf{Different Keyword-to-non-keyword Ratios}} \\
\cmidrule(lr){3-7}
& & 1:4 & 1:5 & 1:6 & 1:7 & 1:8 \\
\midrule
\multirow{7}{*}{\textbf{ESC-50}}
& Unadapted & 61.75 / 85.97 & 61.92 / 88.01 & 62.40 / 89.52 & 62.07 / 90.52 & 61.87 / 91.32 \\
& TBN       & 74.67\pmS{0.05} / 88.26\pmS{0.04} & 73.64\pmS{0.09} / 89.07\pmS{0.03} & 72.54\pmS{0.13} / 89.60\pmS{0.08} & 70.87\pmS{0.03} / 89.78\pmS{0.03} & 69.22\pmS{0.10} / 89.86\pmS{0.05} \\
& Tent      & 74.59\pmS{0.10} / 88.29\pmS{0.05} & 73.44\pmS{0.13} / 89.07\pmS{0.06} & 72.54\pmS{0.32} / 89.71\pmS{0.12} & 70.83\pmS{0.19} / 89.82\pmS{0.08} & 69.15\pmS{0.20} / 89.88\pmS{0.12} \\
& SAR       & 74.55\pmS{0.21} / 88.28\pmS{0.09} & 73.56\pmS{0.26} / 89.11\pmS{0.09} & 72.26\pmS{0.24} / 89.52\pmS{0.07} & 70.65\pmS{0.09} / 89.70\pmS{0.08} & 69.16\pmS{0.16} / 89.88\pmS{0.07} \\
& ETA       & 74.48\pmS{0.12} / 88.19\pmS{0.09} & 73.50\pmS{0.21} / 89.08\pmS{0.08} & 72.42\pmS{0.25} / 89.57\pmS{0.06} & 70.69\pmS{0.17} / 89.76\pmS{0.03} & 69.25\pmS{0.07} / 89.89\pmS{0.03} \\
& AdaKWS    & 74.58\pmS{0.18} / 88.35\pmS{0.06} & 73.50\pmS{0.29} / 89.17\pmS{0.13} & 72.47\pmS{0.14} / 89.68\pmS{0.10} & 71.02\pmS{0.14} / 89.93\pmS{0.04} & 69.31\pmS{0.37} / 90.10\pmS{0.17} \\
& ImKWS     & \textbf{75.24\pmS{0.13} / 88.89\pmS{0.09}} & \textbf{74.31\pmS{0.18} / 89.86\pmS{0.11}} & \textbf{73.58\pmS{0.28} / 90.51\pmS{0.18}} & \textbf{72.00\pmS{0.08} / 90.79\pmS{0.07}} & \textbf{70.76\pmS{0.14} / 91.07\pmS{0.12}} \\
\midrule
\multirow{7}{*}{\textbf{MS-SNSD}}
& Unadapted & 61.83 / 85.75 & 61.96 / 87.73 & 61.86 / 89.05 & 61.59 / 90.01 & 61.33 / 90.75 \\
& TBN       & 70.95\pmS{0.08} / 87.15\pmS{0.03} & 70.28\pmS{0.08} / 88.24\pmS{0.03} & 69.62\pmS{0.08} / 89.01\pmS{0.03} & 67.85\pmS{0.09} / 89.27\pmS{0.02} & 66.68\pmS{0.12} / 89.53\pmS{0.07} \\
& Tent      & 70.70\pmS{0.20} / 87.23\pmS{0.09} & 70.29\pmS{0.44} / 88.43\pmS{0.19} & 69.29\pmS{0.24} / 89.07\pmS{0.10} & 68.00\pmS{0.20} / 89.50\pmS{0.02} & 66.97\pmS{0.16} / 89.87\pmS{0.12} \\
& SAR       & 70.45\pmS{0.05} / 87.04\pmS{0.05} & 69.93\pmS{0.21} / 88.20\pmS{0.05} & 69.36\pmS{0.23} / 88.99\pmS{0.11} & 67.63\pmS{0.10} / 89.25\pmS{0.05} & 66.51\pmS{0.25} / 89.54\pmS{0.10} \\
& ETA       & 70.59\pmS{0.20} / 87.06\pmS{0.12} & 69.95\pmS{0.14} / 88.21\pmS{0.04} & 69.26\pmS{0.36} / 88.98\pmS{0.20} & 67.49\pmS{0.24} / 89.25\pmS{0.06} & 66.48\pmS{0.16} / 89.59\pmS{0.01} \\
& AdaKWS    & 70.66\pmS{0.14} / 87.31\pmS{0.10} & 70.33\pmS{0.11} / 88.56\pmS{0.06} & 69.49\pmS{0.22} / 89.28\pmS{0.12} & 68.01\pmS{0.05} / 89.67\pmS{0.03} & 67.09\pmS{0.12} / 90.04\pmS{0.08} \\
& ImKWS     & \textbf{71.33\pmS{0.05} / 88.13\pmS{0.07}} & \textbf{71.28\pmS{0.41} / 89.59\pmS{0.15}} & \textbf{71.27\pmS{0.19} / 90.66\pmS{0.07}} & \textbf{69.87\pmS{0.02} / 91.13\pmS{0.04}} & \textbf{69.73\pmS{0.16} / 91.74\pmS{0.07}} \\
\bottomrule
\end{tabular}}
\vspace{-12pt}
\end{table*}

% -------- Table 3: Ablation --------
\begin{table}[t]
\centering
\caption{Ablation study on ESC-50 and MS-SNSD environmental noise (1:8) at -10 dB.}
\vspace{-2mm}
\label{tab:4}
\setlength{\tabcolsep}{10pt}
\renewcommand{\arraystretch}{0.9}
\scalebox{0.8}{
\begin{tabular}{lcc|cc}
\toprule
\multirow{2.5}{*}{\textbf{Methods}} & \multicolumn{2}{c}{\textbf{ESC-50}}  & \multicolumn{2}{|c}{\textbf{MS-SNSD}}\\
\cmidrule(lr){2-5}
& ma F1 $\uparrow$  & mi F1 $\uparrow$  & ma F1 $\uparrow$  & mi F1 $\uparrow$ \\
\midrule
ImKWS & \textbf{70.91} & \textbf{91.20} & \textbf{69.91} & \textbf{91.82}\\
w/o DEM     & 70.24 & 90.62 & 68.39 & 90.75\\
w/o Consistency & 69.96 & 90.53 & 69.06 & 91.15\\
w/o Selection & 70.19 & 90.62 & 68.00 & 90.56\\
\bottomrule
\end{tabular}}
\vspace{-12pt}
\end{table}

\section{Results}

\subsection{Effect of Adaptation under Noise with Different SNRs}
Under the severely imbalanced setting (ratio = 1:8) with different SNRs, ImKWS actively suppresses this majority-class overconfidence, as illustrated in Table 1. Compared to the most reliable baseline, AdaKWS, our method improves Macro F1 by +1.23\% / +1.43\% / +1.62\% on ESC-50 at -10 / 0 / 10 dB. The margin is even wider on MS-SNSD, showing improvements of +2.96\% / +2.19\% / +1.50\%, with the most significant gains occurring under the highly challenging -10 dB condition. Crucially, this surge in Macro F1 is accompanied by consistent improvements in Micro F1 (up to +1.87\% on MS-SNSD at -10 dB). This simultaneous improvement strongly indicates that the decoupled penalty branch effectively mitigates the precision-recall trade-off: it maintains sensitivity to rare keywords without excessively inflating false positive predictions on the background class. Ultimately, these results validate that ImKWS prevents the severe gradient fluctuations and majority-class collapse that plague standard EM approaches in realistic, streaming KWS environments.

\subsection{Adaptive Robustness of Diverse Imbalance Ratios} 
% \textcolor{red}{[I think we can use this long table as the main table and place it in Table 1, while the short table cannot accommodate the standard deviation.]}

% Table~\ref{tab:3} summarizes the adaptation performance at the most challenging noise condition (-10 dB) under progressively more severe class imbalance, where the keyword-to-non-keyword ratio ranges from 1:4 to 1:8. Across both ESC-50 and MS-SNSD, ImKWS consistently achieves the best results in every imbalance setting. Compared with the strongest baseline AdaKWS, our method yields consistent Macro F1 improvements across all ratios, and the gains become larger as the imbalance becomes more severe; for example, on MS-SNSD the Macro F1 gap increases from 0.65 at 1:4 to 2.67 at 1:8. A similar trend is observed on ESC-50, where ImKWS maintains a clear advantage over all baselines from 1:4 through 1:8. Meanwhile, ImKWS also improves Micro F1 across all ratios, suggesting that the gains are not obtained by sacrificing majority-class performance. Overall, these results demonstrate that ImKWS provides stable and reliable test-time adaptation under low-SNR conditions and a wide range of imbalance levels, effectively mitigating the majority-class bias that commonly degrades EM-based adaptation in realistic KWS streams.Adaptation method: deyo

Table~\ref{tab:3} summarizes the adaptation performance (mean ± standard deviation) at the most challenging noise condition (-10 dB) under progressively more severe class imbalance, ranging from 1:4 to 1:8. Across both ESC-50 and MS-SNSD, ImKWS consistently achieves the best results in every setting. 

Crucially, this table empirically validates the regularizing effect of our decoupled penalty branch. As the non-keyword ratio increases from 1:4 to 1:8, standard EM methods like AdaKWS suffer from an exponentially accumulating gradient bias toward the background class. In contrast, ImKWS strictly throttles this logit explosion. This theoretical advantage is evidenced by the widening performance gap: at a moderate 1:4 ratio on MS-SNSD, ImKWS yields a reliable Macro F1 improvement of +0.67\% over AdaKWS, but at the extreme 1:8 ratio, this margin expands significantly to +2.64\%. 

A similar trend is observed on ESC-50, where ImKWS maintains a consistent, multi-run verified advantage over all baselines from 1:4 through 1:8. Furthermore, ImKWS simultaneously improves Micro F1 across all ratios, proving that the penalty parameter ($\alpha$) does not achieve keyword sensitivity by sacrificing majority-class precision. Overall, these results demonstrate that ImKWS provides stable TTA that scales robustly with the severity of the data imbalance.

\subsection{Ablation Study}
% | **消融 “w/o PLPD”** | 当前只有 w/o DEM、w/o consistency，缺少“仅用熵、不用 PLPD”的对比。 | 加一列 **w/o PLPD**（即只做熵筛选 + 重加权 + consistency，不做 PLPD 过滤）。若 PLPD 带来提升，可强化“两阶段筛选”的叙事。 |
% Table~\ref{tab:ablation} ablates on ESC-50 (1:8), all SNRs at $-10$ dB. Removing decoupled entropy minimization (w/o DEM) or the consistency loss (w/o consist) each reduces accuracy and macro F1, confirming both components contribute.

% Table~\ref{tab:4} shows the results of the ablation study for ImKWS under two types of environmental noise at -10 dB, isolating the contributions of its two key components: Decoupled Entropy Minimization (DEM) and Multi-view Consistency Loss. When all components are included, ImKWS achieves the highest performance on both datasets (ESC-50: 70.78/91.07; MS-SNSD: 69.62/91.63 in macro/micro F1), confirming the complementary roles of these mechanisms. Removing DEM (w/o dem) leads to a clear drop, especially in macro F1 (ESC-50: 70.78$\rightarrow$70.13; MS-SNSD: 69.62$\rightarrow$68.14), indicating that DEM stabilizes entropy-based adaptation under severe imbalance and mitigates majority-class bias. Similarly, removing multi-view consistency loss (w/o consist) also degrades performance (ESC-50: 70.78$\rightarrow$69.96; MS-SNSD: 69.62$\rightarrow$69.06), confirming that consistency regularization improves robustness by enforcing stable predictions and gradients across perturbed views.

% \subsection{Ablation Study and Hyperparameter Sensitivity}

Table~\ref{tab:4} isolates the contributions of the two core components of ImKWS: Decoupled Entropy Minimization (DEM) and Multi-view Consistency Loss, evaluated under the most challenging -10 dB SNR and 1:8 imbalance ratio. 

The full ImKWS model achieves the highest performance across both datasets. Removing DEM (w/o DEM) forces the model to revert to standard EM, resulting in a severe drop in Macro F1 (e.g., from 69.91\% to 68.39\% on MS-SNSD). This confirms that DEM is the primary mechanism preventing the majority-class logit explosion. Similarly, removing the multi-view consistency loss (w/o Consistency) disrupts adaptation stability, leading to performance degradation (from 69.91\% to 69.06\% on MS-SNSD). This confirms that consistency regularization is vital for enforcing stable predictions across perturbed views in low-SNR conditions.

\subsection{Adaptive Robustness and Gradient Stability}
Figure 2 illustrates the per-class F1 dynamics across varying imbalance ratios. A common failure mode of standard EM in test-time adaptation is shifting the decision boundary to favor the majority class, which artificially boosts Non-keyword F1 but causes a severe, unacceptable drop in Keyword detection sensitivity % 
. However, Figure 2 demonstrates that ImKWS consistently improves Keyword F1 while strictly maintaining, and often boosting, Non-keyword F1. This empirically confirms that the decoupled penalty branch actively calibrates the majority-class logits rather than blindly lowering the detection threshold, ensuring reliable keyword sensitivity without introducing excessive background false positives.

Furthermore, Figure 3 visualizes the gradient norm distributions during the adaptation process, directly validating the stabilizing effect of the multi-view consistency loss ($\mathcal{L}_{consist}$). Under severe class imbalance, relying solely on DEM produces erratic, highly varied gradients, evidenced by the long tails and extreme outliers in the boxplots. By enforcing prediction agreement across augmented audio views using Symmetric Cross-Entropy, the consistency constraint effectively bounds the gradient divergence. When the model is highly uncertain and generates disparate logits for perturbed inputs, $\mathcal{L}_{consist}$ acts as a spatial regularizer, preventing the erratic, high-magnitude gradient spikes that typically arise from DEM's penalty relaxation. As shown in the figure, adding $\mathcal{L}_{consist}$ visibly flattens the upper tail of the gradient distributions
. This suppression of abrupt, aggressive parameter updates ensures stable, monotonic adaptation trajectories, even in the extreme 1:8 imbalance scenario.

% \begin{figure}[t!]
%     \centering
%     \begin{minipage}[t]{0.5\columnwidth}
%         \centering
%         \includegraphics[width=\linewidth]{ms-snsd_pf.pdf}
%         \vspace{-4pt}
%     \end{minipage}
%     \hspace{-6pt}
%     \begin{minipage}[t]{0.5\columnwidth}
%         \centering
%         \includegraphics[width=\linewidth]{ms-snsd_nf.pdf}
%         \vspace{-4pt}
%     \end{minipage}
%     \vspace{-20pt}
%     \caption{Keyword F1 and Non-keyword F1 on MS-SNSD nvironmental noise (1:8) at -10 dB.}
%     \label{fig:f1}
%     \vspace{-10pt}
% \end{figure}

\begin{figure}[t!]
    \centering
    \includegraphics[width=0.48\textwidth]{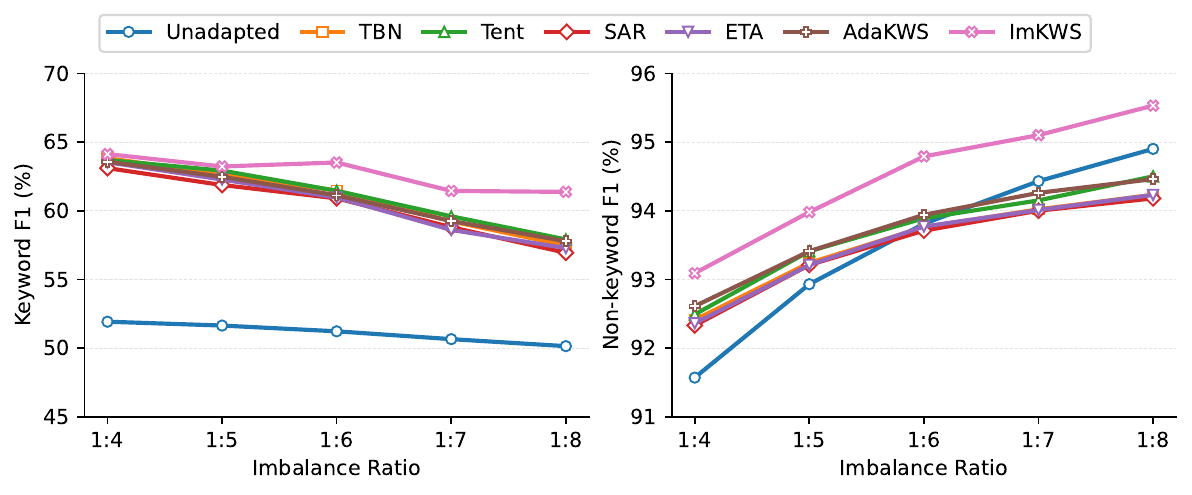}
    \vspace{-20pt}
    \caption{Keyword F1 and Non-keyword F1 on MS-SNSD nvironmental noise at -10 dB.}
    \label{fig:f1}
    \vspace{-10pt}
\end{figure}

\begin{figure}[t!]
    \centering
    \begin{minipage}[t]{0.5\columnwidth}
        \centering
        \includegraphics[width=1.04\linewidth]{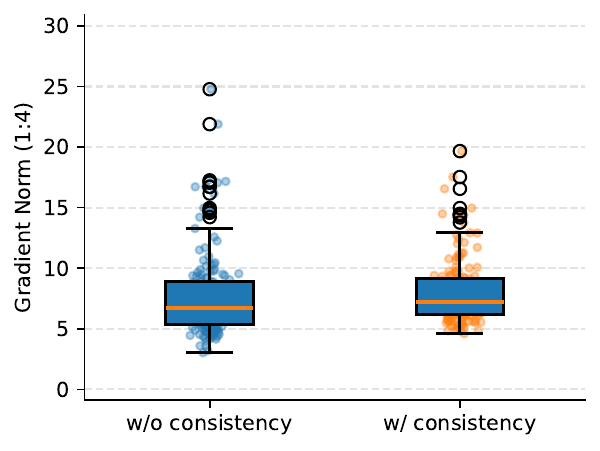}
        \vspace{-4pt}
    \end{minipage}
    \hspace{-4.4pt}
    \begin{minipage}[t]{0.5\columnwidth}
        \centering
        \includegraphics[width=1.04\linewidth]{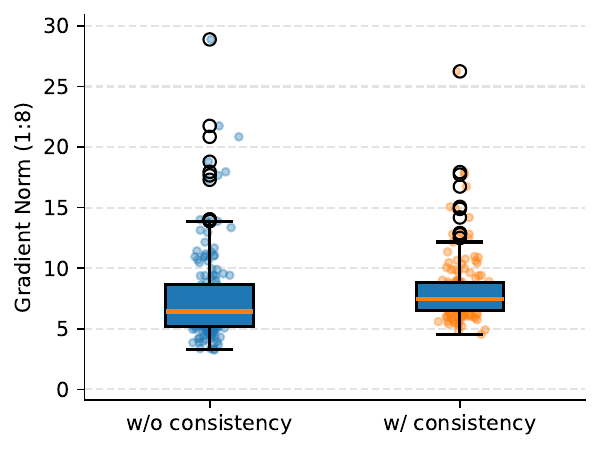}
        \vspace{-4pt}
    \end{minipage}
    \vspace{-20pt}
    \caption{Gradient Norm without and with consistency on MS-SNSD nvironmental noise (1:4 \& 1:8) at -10 dB.}
    \label{fig:norm}
    \vspace{-15pt}
\end{figure}

% \section{Conclusion}

% \textcolor{red}{This work targets test-time adaptation for KWS under realistic extreme keyword and non-keyword imbalance. ImKWS is introduced to mitigate overconfident majority-class drift by decoupling entropy minimization into reward and penalty branches with separately tuned update strengths, giving higher sensitivity to the rare keyword class while remaining conservative on the non-keyword majority. Multi-view augmentation with consistency regularization is further employed to stabilize predictions and gradients under perturbations, improving robustness in dynamic noisy environments. Experiments under diverse conditions show robust adaptation than other baselines.}
\section{Conclusion}
In this paper, we presented ImKWS, a robust test-time adaptation framework tailored for keyword spotting under severe class imbalance and acoustic shifts. By decoupling standard entropy minimization into independent reward and penalty branches, our method effectively resolves the majority-class collapse while preserving keyword sensitivity. Furthermore, incorporating multi-view consistency ensures stable gradient trajectories during streaming adaptation. Extensive evaluations demonstrate that ImKWS significantly outperforms standard adaptation baselines, particularly under extreme imbalance ratios (e.g., 1:8) and low SNR conditions (-10 dB). Future work will explore extending this decoupled framework to memory-constrained on-device learning scenarios.
% \textcolor{red}{[The conclusion has not been revised.]}

\clearpage

\section{Generative AI Use Disclosure}
We use generative AI tools for polishing the manuscript, e.g., correcting the grammar.
% \section{Acknowledgments}

% \ifcameraready
% Acknowledgments (add for camera-ready; omit for review).
% \else
% % [Omitted for double-blind review.]
% \fi

% \clearpage
\bibliographystyle{IEEEtran}
\bibliography{mybib}

@article{lopez2021deep,
  title={Deep spoken keyword spotting: An overview},
  author={L{\'o}pez-Espejo, Iv{\'a}n and Tan, Zheng-Hua and Hansen, John HL and Jensen, Jesper},
  journal={IEEE Access},
  volume={10},
  pages={4169--4199},
  year={2021},
  publisher={IEEE}
}

@article{zhang2017hello,
  title={Hello edge: Keyword spotting on microcontrollers},
  author={Zhang, Yundong and Suda, Naveen and Lai, Liangzhen and Chandra, Vikas},
  journal={arXiv preprint arXiv:1711.07128},
  year={2017}
}

@inproceedings{kim2021broadcasted,
  title={Broadcasted Residual Learning for Efficient Keyword Spotting},
  author={Kim, Byeonggeun and Chang, Simyung and Lee, Jinkyu and Sung, Dooyong},
  booktitle={Proc. Interspeech 2021},
  pages={4538--4542},
  year={2021}
}

@inproceedings{ng2023small,
  title={Small Footprint Multi-channel Network for Keyword Spotting with Centroid Based Awareness},
  author={Ng, Dianwen and Xiao, Yang and Yip, Jia Qi and Yang, Zhao and Tian, Biao and Fu, Qiang and Chng, Eng Siong and Ma, Bin},
  booktitle={Proc. Interspeech 2023},
  pages={296--300},
  year={2023}
}

@inproceedings{xiao2022rainbow,
  title={Rainbow Keywords: Efficient Incremental Learning for Online Spoken Keyword Spotting},
  author={Xiao, Yang and Hou, Nana and Chng, Eng Siong},
  booktitle={Proc. Interspeech 2022},
  pages={3764--3768},
  year={2022}
}

@inproceedings{peng2025dark,
  title={Dark experience for incremental keyword spotting},
  author={Peng, Tianyi and Xiao, Yang},
  booktitle={ICASSP 2025-2025 IEEE International Conference on Acoustics, Speech and Signal Processing (ICASSP)},
  pages={1--5},
  year={2025},
  organization={IEEE}
}

@inproceedings{xiao2025analytickws,
  title={AnalyticKWS: towards exemplar-free analytic class incremental learning for small-footprint keyword spotting},
  author={Xiao, Yang and Tianyi, Peng and Das, Rohan Kumar and Hu, Yuchen and Zhuang, Huiping},
  booktitle={Findings of the Association for Computational Linguistics: ACL 2025},
  pages={14147--14158},
  year={2025}
}

@article{ding2025keyword,
  title={Keyword Mamba: Spoken keyword spotting with state space models},
  author={Ding, Hanyu and Dong, Wenlong and Mao, Qirong},
  journal={Computer Speech \& Language},
  pages={101909},
  year={2025},
  publisher={Elsevier}
}

@inproceedings{wu2020domain,
  title={Domain Aware Training for Far-Field Small-Footprint Keyword Spotting},
  author={Wu, Haiwei and Jia, Yan and Nie, Yuanfei and Li, Ming},
  booktitle={Proc. Interspeech 2020},
  pages={2562--2566},
  year={2020}
}

@article{lopez2021novel,
  title={A novel loss function and training strategy for noise-robust keyword spotting},
  author={L{\'o}pez-Espejo, Iv{\'a}n and Tan, Zheng-Hua and Jensen, Jesper},
  journal={IEEE/ACM Transactions on Audio, Speech, and Language Processing},
  volume={29},
  pages={2254--2266},
  year={2021},
  publisher={IEEE}
}

@inproceedings{yang2025cross,
  title={Cross-domain few-shot open-set keyword spotting using keyword adaptation and prototype reprojection},
  author={Yang, Mingru and He, Qianhua and Huang, Jinxin and Chen, Yongqiang and Liu, Zunxian and Li, Yanxiong},
  booktitle={ICASSP 2025-2025 IEEE International Conference on Acoustics, Speech and Signal Processing (ICASSP)},
  pages={1--5},
  year={2025},
  organization={IEEE}
}

@inproceedings{ozay2024joint,
  title={Joint embedding learning and latent subspace probing for cross-domain few-shot keyword spotting},
  author={Ozay, Mete},
  booktitle={ICASSP 2024-2024 IEEE International Conference on Acoustics, Speech and Signal Processing (ICASSP)},
  pages={6425--6429},
  year={2024},
  organization={IEEE}
}

@inproceedings{ganin2015unsupervised,
  title={Unsupervised domain adaptation by backpropagation},
  author={Ganin, Yaroslav and Lempitsky, Victor},
  booktitle={International conference on machine learning},
  pages={1180--1189},
  year={2015},
  organization={PMLR}
}

@inproceedings{hou2019domain,
  title={Domain adversarial training for improving keyword spotting performance of esl speech},
  author={Hou, Jingyong and Guo, Pengcheng and Sun, Sining and Soong, Frank K and Hu, Wenping and Xie, Lei},
  booktitle={ICASSP 2019-2019 IEEE International conference on acoustics, speech and signal processing (ICASSP)},
  pages={8122--8126},
  year={2019},
  organization={IEEE}
}

@inproceedings{liu2023te,
  title={TE-KWS: text-informed speech enhancement for noise-robust keyword spotting},
  author={Liu, Dong and Mao, Qirong and Gao, Lijian and Ren, Qinghua and Chen, Zhenghan and Dong, Ming},
  booktitle={Proceedings of the 31st ACM International Conference on Multimedia},
  pages={601--610},
  year={2023}
}

@article{lim2023joint,
  title={Joint framework of curriculum learning and knowledge distillation for noise-robust and small-footprint keyword spotting},
  author={Lim, Jaebong and Baek, Yunju},
  journal={IEEE Access},
  volume={11},
  pages={100540--100553},
  year={2023},
  publisher={IEEE}
}

@article{liang2025comprehensive,
  title={A comprehensive survey on test-time adaptation under distribution shifts},
  author={Liang, Jian and He, Ran and Tan, Tieniu},
  journal={International Journal of Computer Vision},
  volume={133},
  number={1},
  pages={31--64},
  year={2025},
  publisher={Springer}
}

@inproceedings{wang2021tent,
  title={Tent: Fully Test-Time Adaptation by Entropy Minimization},
  author={Dequan Wang and Evan Shelhamer and Shaoteng Liu and Bruno Olshausen and Trevor Darrell},
  booktitle={International Conference on Learning Representations},
  year={2021}
}

@inproceedings{niu2023towards,
  title={Towards Stable Test-time Adaptation in Dynamic Wild World}, 
  author={Shuaicheng Niu and Jiaxiang Wu and Yifan Zhang and Zhiquan Wen and Yaofo Chen and Peilin Zhao and Mingkui Tan},
  booktitle={The Eleventh International Conference on Learning Representations },
  year={2023}
}

@inproceedings{niu2022efficient,
  title={Efficient test-time model adaptation without forgetting},
  author={Niu, Shuaicheng and Wu, Jiaxiang and Zhang, Yifan and Chen, Yaofo and Zheng, Shijian and Zhao, Peilin and Tan, Mingkui},
  booktitle={International conference on machine learning},
  pages={16888--16905},
  year={2022},
  organization={PMLR}
}

@inproceedings{schneider2020improving,
  title={Improving robustness against common corruptions by covariate shift adaptation},
  author={Schneider, Steffen and Rusak, Evgenia and Eck, Luisa and Bringmann, Oliver and Brendel, Wieland and Bethge, Matthias},
  booktitle={Proceedings of the 34th International Conference on Neural Information Processing Systems},
  pages={11539--11551},
  year={2020}
}

@inproceedings{lin2022listen,
  title={Listen, Adapt, Better WER: Source-free Single-utterance Test-time Adaptation for Automatic Speech Recognition},
  author={Lin, Guan-Ting and Li, Shang-Wen and Lee, Hung-yi},
  booktitle={Proc. Interspeech 2022},
  pages={2198--2202},
  year={2022}
}

@inproceedings{dong2025ebats,
  title={E-{BATS}: Efficient Backpropagation-Free Test-Time Adaptation for Speech Foundation Models},
  author={Jiaheng Dong and Hong Jia and Soumyajit Chatterjee and Abhirup Ghosh and James Bailey and Ting Dang},
  booktitle={The Thirty-ninth Annual Conference on Neural Information Processing Systems},
  year={2025}
}

@inproceedings{xiao2025adakws,
  title={AdaKWS: Towards Robust Keyword Spotting with Test-Time Adaptation},
  author={Xiao, Yang and Peng, Tianyi and Zhou, Yanghao and Das, Rohan Kumar},
  booktitle={Proc. Interspeech 2025},
  pages={5408--5412},
  year={2025}
}

@article{warden2018speech,
  title={Speech commands: A dataset for limited-vocabulary speech recognition},
  author={Warden, Pete},
  journal={arXiv preprint arXiv:1804.03209},
  year={2018}
}

@inproceedings{piczak2015esc,
  title={ESC: Dataset for environmental sound classification},
  author={Piczak, Karol J},
  booktitle={Proceedings of the 23rd ACM international conference on Multimedia},
  pages={1015--1018},
  year={2015}
}

@inproceedings{reddy2019scalable,
  title={A Scalable Noisy Speech Dataset and Online Subjective Test Framework},
  author={Reddy, Chandan KA and Beyrami, Ebrahim and Pool, Jamie and Cutler, Ross and Srinivasan, Sriram and Gehrke, Johannes},
  booktitle={Proc. Interspeech 2019},
  pages={1816--1820},
  year={2019}
}

@inproceedings{wang2019symmetric,
  title={Symmetric cross entropy for robust learning with noisy labels},
  author={Wang, Yisen and Ma, Xingjun and Chen, Zaiyi and Luo, Yuan and Yi, Jinfeng and Bailey, James},
  booktitle={Proceedings of the IEEE/CVF international conference on computer vision},
  pages={322--330},
  year={2019}
}

@inproceedings{xiao2025dg,
  title={DG-SED: Domain Generalization for Sound Event Detection with Heterogeneous Training Data},
  author={Xiao, Yang and Yin, Han and Bai, Jisheng and Das, Rohan Kumar},
  booktitle={2025 Asia Pacific Signal and Information Processing Association Annual Summit and Conference (APSIPA ASC)},
  pages={143--148},
  year={2025},
  organization={IEEE}
}

@article{xiao2026adapting,
  title={Adapting Where It Matters: Depth-Aware Adaptation for Efficient Multilingual Speech Recognition in Low-Resource Languages},
  author={Xiao, Yang and Holden, Eun-Jung and Dang, Ting},
  journal={arXiv preprint arXiv:2602.01008},
  year={2026}
}

@article{luo2026survey,
  title={A Survey of Large Audio Language Models: Generalization, Trustworthiness, and Outlook},
  author={Luo, Kaiwen and Zhou, Zhenhong and Wang, Leo and Lin, Liang and Xiao, Yang and Shao, Tianyu and Zhang, Yuanhe and Li, Yuxuan and Yu, Miao and Lyu, Kailin and others},
  journal={arXiv preprint arXiv:2605.20266},
  year={2026}
}

@inproceedings{dong2026test,
  title={Test-Time Adaptation for Speech Emotion Recognition},
  author={Dong, Jiaheng and Jia, Hong and Dang, Ting},
  booktitle={ICASSP 2026-2026 IEEE International Conference on Acoustics, Speech and Signal Processing (ICASSP)},
  pages={17157--17161},
  year={2026},
  organization={IEEE}
}

@inproceedings{dang2026test,
  title={Test-Time Scaling for Auditory Cognition in Audio Language Models},
  author={Dang, Ting and Gao, Yan and Jia, Hong},
  booktitle={ICASSP 2026-2026 IEEE International Conference on Acoustics, Speech and Signal Processing (ICASSP)},
  pages={16582--16586},
  year={2026},
  organization={IEEE}
}

@inproceedings{shi2026tta,
 title={Emo-TTA: Improving Test-Time Adaptation of Audio-Language Models for Speech Emotion Recognition},
 author={Shi, Jiacheng and Du, Hongfei and Hong, Y Alicia and Gao, Ye},
 booktitle={ICASSP 2026-2026 IEEE International Conference on Acoustics, Speech and Signal Processing (ICASSP)},
 pages={16412--16416},
 year={2026},
 organization={IEEE}
}

\end{document}